# Surface Transport in the $\nu$=0 Quantum Hall Ferromagnetic State in the Organic Dirac Fermion System

Toshihito Osada*

*Institute for Solid State Physics, University of Tokyo,*

*5-1-5 Kashiwanoha, Kashiwa, Chiba 277-8581, Japan.*

We discuss the surface magnetotransport in the quantum Hall (QH) ferromagnetic state expected in the organic Dirac fermion system $\alpha$-(BEDT-TTF)$_2$I$_3$. The QH ferromagnetic state is one of the possible $\nu$=0 QH states in the two-dimensional Dirac fermion system resulting from the degeneracy breaking of the $n$=0 Landau level. It is characterized by the helical edge state. We have studied the interlayer surface transport via helical edge state in the multilayer QH ferromagnet, in which the bulk region is insulating. We have clarified that the surface conductivity is much less than $e^2/h$ and decreases as the magnetic field is tilted to the normal direction of the side surface. These features explain the observed interlayer magnetoresistance in $\alpha$-(BEDT-TTF)$_2$I$_3$.

A layered organic conductor $\alpha$-(BEDT-TTF)$_2$I$_3$, where BEDT-TTF denotes bis(ethylenedithio)-tetrathiafulvalene, has attracted a great deal attention since it was revealed to be a two-dimensional (2D) massless Dirac fermion system like graphene under high pressure[1,2]. Since the coupling between BEDT-TTF conducting layers is very small (interlayer transfer energy, $t_c$, is much less than 1 meV), this compound is usually regarded as a 2D system. At ambient pressure, it undergoes a phase transition into the charge-ordered insulating phase at $T = 135$ K. Under high pressure ($P > 1.5$ GPa), this transition is suppressed so that the metallic phase survives down to low temperatures. According to the tight-binding band calculation in the metallic phase, each BEDT-TTF layer has 2D band dispersion in which the conduction band and the valence band contact forming a pair of Dirac cones[3,4]. In contrast to graphene, the Dirac cones are tilted and anisotropic, and they exist at two general points $\mathbf{k}_0$ and $-\mathbf{k}_0$ (called valleys) in the 2D Brillouin zone. The Femi level is fixed at the Dirac point because of the crystal stoichiometry. $\alpha$-(BEDT-TTF)$_2$I$_3$ was the first example of the 2D massless Dirac fermion system in the bulk crystal.

One of the most characteristic features of 2D massless Dirac fermions under the magnetic field is the anomalous $n=0$ Landau level (LL), of which energy is always equal to the Dirac point energy. Four states with different spin (up or down) and valley ($\mathbf{k}_0$ or $-\mathbf{k}_0$) are degenerate in the $n=0$ LL. It has a finite broadening width, $\Gamma$, due to scattering and interlayer tunneling. The quantum limit, where the Fermi level is located only in the $n=0$ LL with no overlap with other LLs, is easily reached even at low magnetic fields, since the cyclotron energy becomes large around the Dirac point reflecting the massless nature. In $\alpha$-(BEDT-TTF)$_2$I$_3$, we can expect the quantum limit above 0.2 T.

It had been known that α-(BEDT-TTF)$_2$I$_3$ shows anomalous transport properties in this field region. The interlayer magnetoresistance (MR) shows remarkable decrease as the magnetic field increases, namely, strong negative MR[5]. Interlayer Hall resistance, which is measured by the interlayer voltage under the in-plane current flow, depends not on the magnetic field strength but only on the magnetic field direction with unusual cot$\theta$ dependence, where $\theta$ is the elevation angle of the magnetic field[6]. These phenomena have been well explained as the electronic properties of the multilayer massless Dirac fermion system at the quantum limit where spin splitting is ignored[7, 8]. In the model, the interlayer coupling was treated as a perturbation following the previous works on interlayer magnetotransport[9]. The negative MR results from the linear increase of the degeneracy of the $n$=0 LL. The agreement between theory and experiments strongly supports the realization of massless Dirac fermions in α-(BEDT-TTF)$_2$I$_3$. It was also confirmed by the specific heat experiment[10].

When spin splitting becomes comparable to the width of the $n$=0 LL in higher magnetic fields, the MR turns to positive because the DOS at $E_F$ begins to decrease. It exponentially increases, approximately obeying $R \sim 1/\exp(-\mu_B B/k_B T)$ at $T > 1\text{K}$, which means activated transport[11]. However, this increase tends to saturate at higher magnetic fields[11].

In this paper, we show that the saturation of interlayer MR in α-(BEDT-TTF)$_2$I$_3$ results from the interlayer surface transport via the helical edge state in the quantum Hall ferromagnetic state, which is one of the possible $\nu$=0 quantum Hall (QH) states. The observed activated behavior is an evidence for the existence of a mobility gap at the Fermi level, in which electronic states cannot contribute to dissipative transport. This fact strongly suggests that the $\nu$=0 QH state is realized as a result of the degeneracy breaking of the $n$=0 LL. In fact, the $\nu \neq 0$ QH effects have been observed in the doped $\alpha$-(BEDT-TTF)$_2$I$_3$ thin film[12]. The saturation after the activated behavior suggests the existence of transport channels other than the insulating bulk channel which shows the activated transport. We consider the scenario that the surface transport channel due to the edge state in the $\nu$=0 QH state limits the interlayer MR causing the saturation[13].

In the 2D massless Dirac fermion system, generally, two types of the $\nu$=0 QH states are possible[14]. One is the QH insulator which appear in the case that the valley splitting is dominant in the degeneracy breaking. It is fully insulating with no edge state, and considered as the ground state of undoped graphene under magnetic fields[15, 16]. Another is the QH ferromagnet which is realized when the spin splitting is dominant in the degeneracy breaking. It is a spin-polarized state and accompanied by a helical edge state, which consists of a pair of edge states with opposite spin and chirality (Fig. 1(a))[17, 18]. This is analogous with the 2D quantum spin Hall system[19]. Although the Chern number is zero in the QH ferromagnet, the spin Chern number, which is defined as the half of difference between Chern numbers of up-spin and down-spin electrons, is one. As long as the spin component along the magnetic field is a good quantum number, the edge state of the QH ferromagnet is topologically protected. In multilayer systems, the helical edge states on 2D layers couple with each other by the interlayer coupling, forming the surface state surrounding the sample side as shown in Fig. 1(b). This surface state could contribute to the interlayer transport.

In below, we discuss interlayer surface transport in the QH ferromagnet. We consider a model of the multilayer Dirac fermion system, where 2D massless Dirac fermion layers parallel to the $xy$-plane stack along the $z$-axis with weak interlayer coupling. The band dispersion of each layer has two isotropic Dirac cones at $\mathbf{k}_0$ and $-\mathbf{k}_0$ in 2D $\mathbf{k}$-space. Each layer located at $z = z_i$ is a semi-finite plane in $x < 0$ with an edge at $x = 0$. The magnetic field $\mathbf{B} = \nabla \times \mathbf{A} = (B_x, B_y, B_z)$ with the general orientation is applied to the system ($B_z > 0$). The effective Hamiltonian $\widehat{H} = \widehat{H}_0 + \widehat{H}'$ is written as the direct sum of those for two valleys:

$$\widehat{H}_0 = \hbar v(\hat{k}_x \sigma_x + \hat{k}_y \sigma_y) \oplus \hbar v(\hat{k}_x \sigma_x - \hat{k}_y \sigma_y) + s\mu_B |\mathbf{B}| I_4 \qquad (1)$$

$$\widehat{H}' = -2t_c \cos c k_z I_4, \qquad (2)$$

where $\hat{\mathbf{k}} = (\hat{k}_x, \hat{k}_y, \hat{k}_z) = -i\nabla + (e/\hbar)\mathbf{A}$. We choose a gauge $\mathbf{A} = (B_y z, B_z x - B_x z, 0)$. $(\sigma_x, \sigma_y, \sigma_z)$ are Pauli matrices, and $I_4$ is a $4 \times 4$ unit matrix. The electron spin in the field direction is indicated by $s = \pm 1$. $v$, $\mu_B$, $t_c$, and $c$ are the velocity of 2D Dirac fermions, the Bohr magneton, the interlayer transfer energy, and the interlayer spacing, respectively. In order to discuss the QH ferromagnet, we do not introduce any inter-valley coupling in $\widehat{H}_0$, so that no valley splitting occurs in the bulk LLs. However, the boundary condition at the edge could cause inter-valley mixing. The electronic state is represented by a four-component spinor, $\mathbf{F}(\mathbf{r}) = {}^t\{F_A^{(\mathbf{k}_0)}(\mathbf{r}), F_B^{(\mathbf{k}_0)}(\mathbf{r}), F_A^{(-\mathbf{k}_0)}(\mathbf{r}), F_B^{(-\mathbf{k}_0)}(\mathbf{r})\}$, where $A$ and $B$ are pseudo-spin indices. One of the typical boundary conditions is given by $F_A^{(\mathbf{k}_0)}(\mathbf{r})|_{x=0} + F_A^{(-\mathbf{k}_0)}(\mathbf{r})|_{x=0} = 0$ and $F_B^{(\mathbf{k}_0)}(\mathbf{r})|_{x=0} + F_B^{(-\mathbf{k}_0)}(\mathbf{r})|_{x=0} = 0$, which corresponds to the armchair edge of graphene[20]. In the following, we employ this boundary condition, since it gives no edge state at zero magnetic field. However, the results do not depend on details of boundary condition.

The eigen states of unperturbed Hamiltonian $\widehat{H}_0$ are edge states of LLs of the 2D Dirac fermion system on each layer. They are labeled by the Landau index $n(= 0, \pm 1, \pm 2, ...)$, the valley splitting index $\tau(= \pm 1)$, the center coordinate $x_0$, the layer position $z_i$, and the real spin index $s(= \pm 1)$. Their four component envelope function are written as $\mathbf{F}_{n\tau x_0 z_i}(\mathbf{r}) = \mathbf{f}_{n\tau x_0}(x)\exp\{i(-eB_y z_i/\hbar)x\}\exp(iK_y y)\delta_{z,z_i}$, where $\mathbf{f}_{n\tau x_0}(x)$ is a four-component function which varies in the range of the magnetic length $l = \sqrt{\hbar/eB_z}$ from the edge. The edge state wave number $K_y$ and the center coordinate $x_0$ relate with $x_0 = -l^2 K_y + (B_x/B_z)z_i$. Figure 2 shows the schematic dispersion of the $n$=0 edge states as a function of $x_0$. Around the sample edge, the up-spin and down-spin $n$=0 LLs split into two branches ($\tau = \pm 1$) with opposite group velocity. At the Fermi energy $E_F$, there are two edge states with opposite spin and current direction on each layer at $z_i$.

To obtain the interlayer surface conductivity $\sigma_{zz}^{(\text{surface})}$, we treat the interlayer coupling $\widehat{H}'$ as a perturbation, and evaluate its lowest order contribution to $\sigma_{zz}^{(\text{surface})}$[9]. Since $\widehat{H}'$ conserves the real spin, the perturbation matrix elements are given by

$$\langle \mathbf{F}_{\alpha' x_0' z_i'}|\widehat{H}'|\mathbf{F}_{\alpha x_0 z_i}\rangle$$

$$= -t_c \left\{ \int_{-\infty}^{0} \mathbf{f}_{\alpha, x_0 + \frac{B_x}{B_z}c}(x)^* \cdot \mathbf{f}_{\alpha, x_0}(x) \exp\left(i\frac{cB_y}{lB_z}\frac{x}{l}\right)dx \right\} \delta_{\alpha',\alpha}\delta_{x_0', x_0 + \frac{B_x}{B_z}c}\delta_{z_i', z_i + c}$$

$$- t_c \left\{ \int_{-\infty}^{0} \mathbf{f}_{\alpha, x_0 - \frac{B_x}{B_z}c}(x)^* \cdot \mathbf{f}_{\alpha, x_0}(x) \exp\left(-i\frac{cB_y}{lB_z}\frac{x}{l}\right)dx \right\} \delta_{\alpha',\alpha}\delta_{x_0', x_0 - \frac{B_x}{B_z}c}\delta_{z_i', z_i - c}.$$

(3)

Here, $\alpha = (n, \tau, s)$ is the index labeling the edge state in Fig. 2. This matrix element leads two selection rules: Tunneling is only allowed between neighboring layers ($z_i' = z_i \pm c$), and the center coordinates of the initial and final states satisfy $x_0' = x_0 \pm$ $(B_x/B_z)c$. Since the energy of edge states depends on $x_0$, the energy generally changes after interlayer tunneling corresponding to the shift of $x_0$ as shown in Fig. 2(b). Therefore, interlayer tunneling occurs with no energy change only when the magnetic field is parallel to the edge surface ($B_x/B_z = 0$) as shown in Fig. 2(a). This causes the resonant inter-edge tunneling at $B_x/B_z = 0$, when the field orientation is changed. The helical surface state is stabilized by this resonant tunneling.

The complex interlayer surface conductivity $\tilde{\sigma}_{zz}^{(\text{surface})}$ can be evaluated by using Kubo formula. $\tilde{\sigma}_{zz}^{(\text{surface})}$ is expanded to a power series of the interlayer coupling $t_c$, and the lowest order contribution of $t_c$ is given by

$$\tilde{\sigma}_{zz}^{(\text{surface})}$$
$$= -\frac{i\hbar}{L_y L_z} \sum_{\alpha,\alpha'} \sum_{x_0,x_0'} \sum_{z_i,z_{i'}} \left|\left\langle \mathbf{F}_{\alpha' x_0' z_{i'}} \left| \hat{j}_z^{(\text{surface})} \right| \mathbf{F}_{\alpha x_0 z_i} \right\rangle\right|^2 \frac{f(E_{\alpha x_0}) - f(E_{\alpha' x_0'})}{E_{\alpha' x_0'} - E_{\alpha x_0}} \frac{1}{E_{\alpha' x_0'} - E_{\alpha x_0} - \frac{i\hbar}{\tau^{(\text{edge})}}}.$$

(4)

Here, $\hat{j}_z^{(\text{surface})} = -ev_z = -(e/i\hbar)[\hat{z}, \widehat{H}']$ is interlayer surface current density, and its matrix element is given by $\left\langle \mathbf{F}_{\alpha' x_0' z_{i'}} \left| \hat{j}_z^{(\text{surface})} \right| \mathbf{F}_{\alpha x_0 z_i} \right\rangle = i(e/\hbar)(z_i' - z_i)\left\langle \mathbf{F}_{\alpha' x_0' z_{i'}} \left| \widehat{H}' \right| \mathbf{F}_{\alpha x_0 z_i} \right\rangle$. In the case of $(B_x^2 + B_y^2)^{1/2}/B_z \ll l/c$, which is satisfied in most field directions, the surface conductivity $\sigma_{zz}^{(\text{surface})} = \text{Re}\left\{\tilde{\sigma}_{zz}^{(\text{surface})}\right\}$ can be easily obtained as

$$\sigma_{zz}^{(\text{surface})} = \frac{4t_c^2 c \tau^{(\text{edge})}}{\hbar^2 v_F^{(\text{edge})}} \left(\frac{e^2}{h}\right) \frac{1}{1 + (B_x/B_x^{(0)})^2}.$$

(5)

Here, $B_x^{(0)} = (h/e)/(2\pi c v_F^{(\text{edge})} \tau^{(\text{edge})})$ and $v_F^{(\text{edge})} = (l^2/\hbar)|dE_{\alpha x_0}/dx_0|_{x_0 = x_{0F}^{(\alpha)}}$.

$v_F^{(\text{edge})}$ is the scattering lifetime of the edge states. $x_{0F}^{(\alpha)}$ is the center coordinate of edge subband $\alpha$ at the Fermi level. We have assumed $k_B T \ll \hbar/\tau^{(\text{bulk})}$ (low temperature) and $t_c \ll \hbar/\tau^{(\text{bulk})}$ (the dirty limit), where $\hbar/\tau^{(\text{bulk})}$ is the scattering width of the $n=0$ LL.

When the field is parallel to the stacking axis ($z$-axis), $\sigma_{zz}^{(\text{surface})}$ shows weak field dependence reflecting that of $v_F^{(\text{edge})}$. The value of $\sigma_{zz}^{(\text{surface})}$ is finite but much less than $e^2/h$ at the limit of $T = 0$, since $v_F^{(\text{edge})} \sim v$, $2t_c c/\hbar \ll v$, and $t_c \ll \hbar/\tau^{(\text{edge})}$. This means that the interlayer surface transport is metallic but diffusive. This is a remarkable feature of the helical surface state as a 2D electron system. Once the field is tilted from the stacking axis to the normal of the edge surface ($x$-axis), $\sigma_{zz}^{(\text{surface})}$ shows the Lorentzian decay due to the inhibition of interlayer tunneling. On the other hand, when the field is tilted in the edge surface ($yz$-plane), $\sigma_{zz}^{(\text{surface})}$ shows no explicit angle-dependence. Therefore, $\sigma_{zz}^{(\text{surface})}$ shows the resonant peak due to the inter-edge resonant tunneling, when the magnetic field is parallel to the edge surface. In real samples, the edge surfaces face various directions, so that the total surface conductance takes the maximum value around the field direction parallel to the stacking axis. These features are expected from the analogy with the chiral surface state in the $\nu \neq 0$ QH multilayer systems[21-23].

In the QH ferromagnet, there still remains the activated bulk transport in addition to the surface transport. So, the interlayer resistance is given by $R_{zz} = 1/(\sigma_{zz}^{(\text{bulk})} S/L_z + \sum \sigma_{zz}^{(\text{surface})} L^{(\text{edge})}/L_z)$, where $S$, $L^{(\text{edge})}$, and $L_z$ are the sectional area, the length of each edge, and the thickness of the plate-like crystal, respectively. The summation is taken for all side (edge) surfaces surrounding the crystal. The interlayer bulk conductivity $\sigma_{zz}^{(bulk)}$ is given in Ref. 7. At low magnetic fields, $\sigma_{zz}^{(bulk)}$ linearly increases with $B_z$ (negative MR), then it exponentially decreases at high fields due to spin splitting[7]. Because the behavior of $R_{zz}$ in tilted magnetic fields depends on the configuration of side surfaces in the crystal, we assume that the contribution of special side surfaces normal to the $x$-axis is dominant.

Figure 3 shows the simulated interlayer resistance $R_{zz}$ under the magnetic fields parallel to the $z$-axis for several temperatures. $R_{zz}$ is normalized by $R_0 \equiv (\hbar^2 c L_z / 4 t_c^2 \tau^{(bulk)2} S)/(e^2/h)$. We can see that the exponential increase of the bulk resistance is limited by the surface conduction. Figure 4 shows the simulated angle-dependence of $R_{zz}$. The distance and the direction from the origin indicate the field strength $|\mathbf{B}|$ and the field orientation, respectively. In the case that the magnetic field is tilted in the $xz$-plane perpendicular to the surface (Fig. 4(a)), $R_{zz}$ shows the saturating behavior only when the field is parallel to the stacking axis ($B_x/B_z = 0$). Once the field is tilted ($B_x \neq 0$), $R_{zz}$ increases monotonously without saturation. Therefore, a resonant dip structure appears at $B_x/B_z = 0$ in the angle-dependence of $R_{zz}$. On the other hand, when the field is tilted in the $yz$-plane parallel to the surface (Fig. 4(b)), $R_{zz}$ shows the saturation for all field directions since $B_x = 0$.

Finally, we compare the above results with the experiments in $\alpha$-(BEDT-TTF)$_2$I$_3$. The behaviors of interlayer MR observed in $\alpha$-(BEDT-TTF)$_2$I$_3$[11] are well reproduced in Fig. 3. The saturated resistance ($\sim 500$ k$\Omega$) and the sample size ($0.5 \times 0.8 \times 0.05$ mm$^3$)[11] give the surface conductivity $\sigma_{zz}^{(surface)} \sim 0.001 e^2/h$ much less than $e^2/h$ as expected. Recently, we have experimentally confirmed the surface transport in $α$-(BEDT-TTF)$_2$I$_3$ by observing that interlayer MR is not scaled by sectional area in the saturation region[24]. Moreover, we have observed that the saturated value became minimum when magnetic field was parallel to the stacking direction[24]. Since these facts are well explained by the present model, the realization of the QH ferromagnet with the helical edge state is strongly suggested in $α$-(BEDT-TTF)$_2$I$_3$.

In conclusion, we have considered the interlayer surface transport caused by the helical surface state in the multilayer QH ferromagnet. Electron tunneling between edge states on two neighboring layers is allowed only when the magnetic field is parallel to the stacking direction. This resonant tunneling causes the resonant increase of interlayer surface conductivity around the normal field orientation. The experimentally observed features of interlayer MR in the organic Dirac fermion system $α$-(BEDT-TTF)$_2$I$_3$ are well explained by considering the surface transport in the multilayer QH ferromagnet.

**Acknowledgements**    The author thanks Prof. N. Tajima, Prof. K. Kajita, Prof. K. Kanoda, Prof. Masatoshi Sato, Prof. Y. Tanaka, and Prof. H. Fukuyama for valuable discussions and comments. He also thanks Dr. Mitsuyuki Sato, Dr. K. Uchida, and Dr. T. Konoike for the collaboration in experiments on this subject. This work was supported by the "Science of Atomic Layers" (No. 25107003) and "Topological Quantum Phenomena" (No. 22103002) Grant-in Aid for Scientific Research on Innovative Areas from the Ministry of Education, Culture, Sports, Science and Technology (MEXT) of Japan.

* osada@issp.u-tokyo.ac.jp

**Figure 1** (T. Osada)

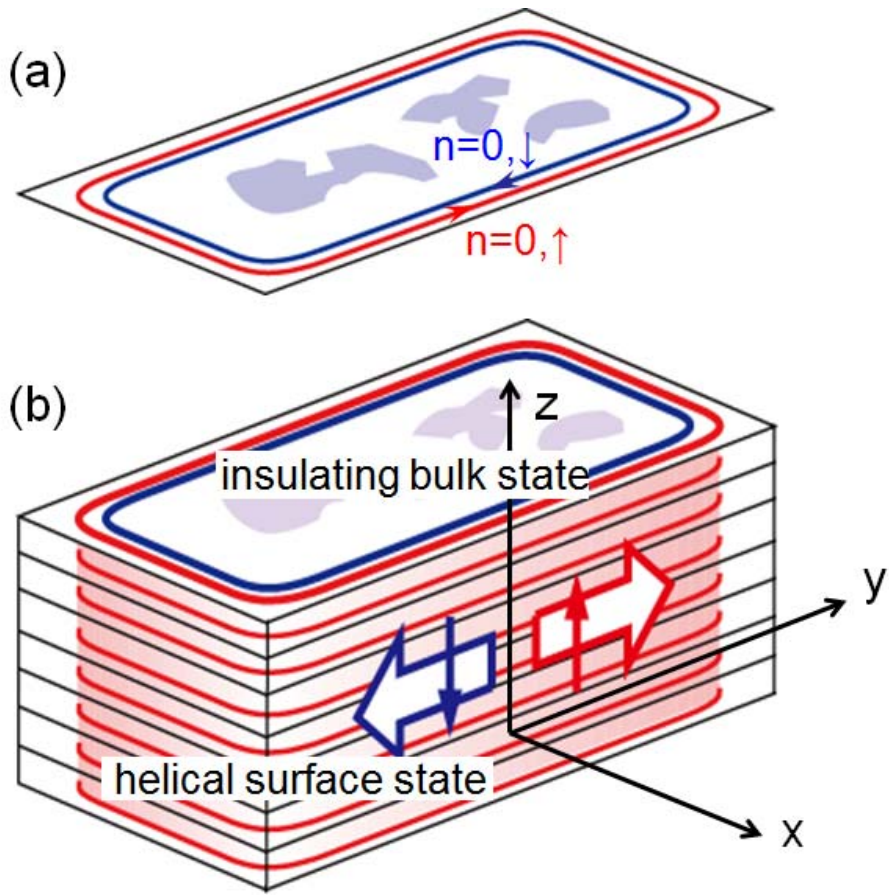

**Fig. 1.** (color online)

Schematic configuration of the helical edge state in the $\nu=0$ QH ferromagnetic state (a) in the 2D Dirac fermion system and (b) in the multilayer Dirac fermion system. The coordinate axes for the front surface are indicated.

**Figure 2** (T. Osada)

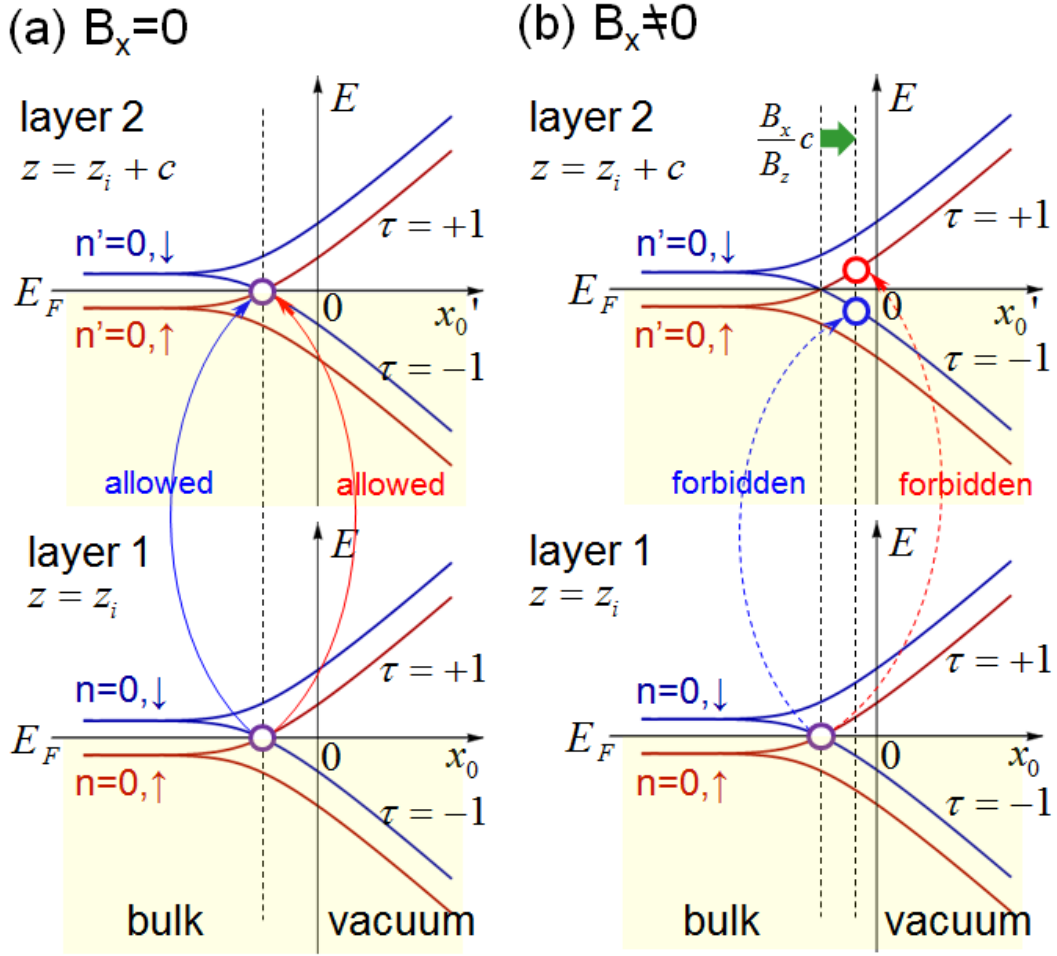

**Fig. 2.** (color online)

Resonant tunneling between helical edge channels on neighboring two layers. (a) Resonant interlayer tunneling when the magnetic field is parallel to the surface. (b) Tunneling conserving energy is not allowed when $B_x$ is finite.

**Figure 3** (T. Osada)

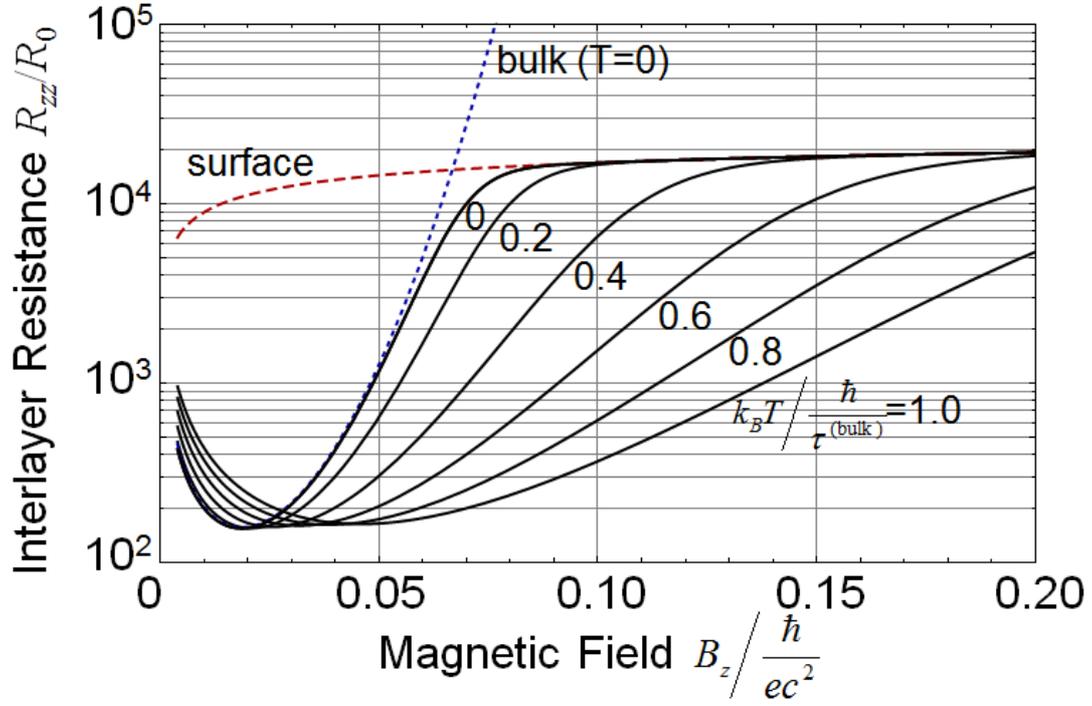

**Fig. 3.** (color online)

Interlayer resistance $R_{zz}$ of the multilayer QH ferromagnet under normal magnetic fields at several temperatures. The dotted curve and dashed curve indicate contributions of bulk transport at $T=0$ and surface transport, respectively. $c = 1.75$ nm, $v = 2.4 \times 10^4$ m/s, $\tau^{(\text{bulk})} = 2$ ps, $\tau^{(\text{edge})} = 20$ ps, and $cL^{(\text{edge})}/S = 10^{-4}$ were assumed.

**Figure 4** (T. Osada)

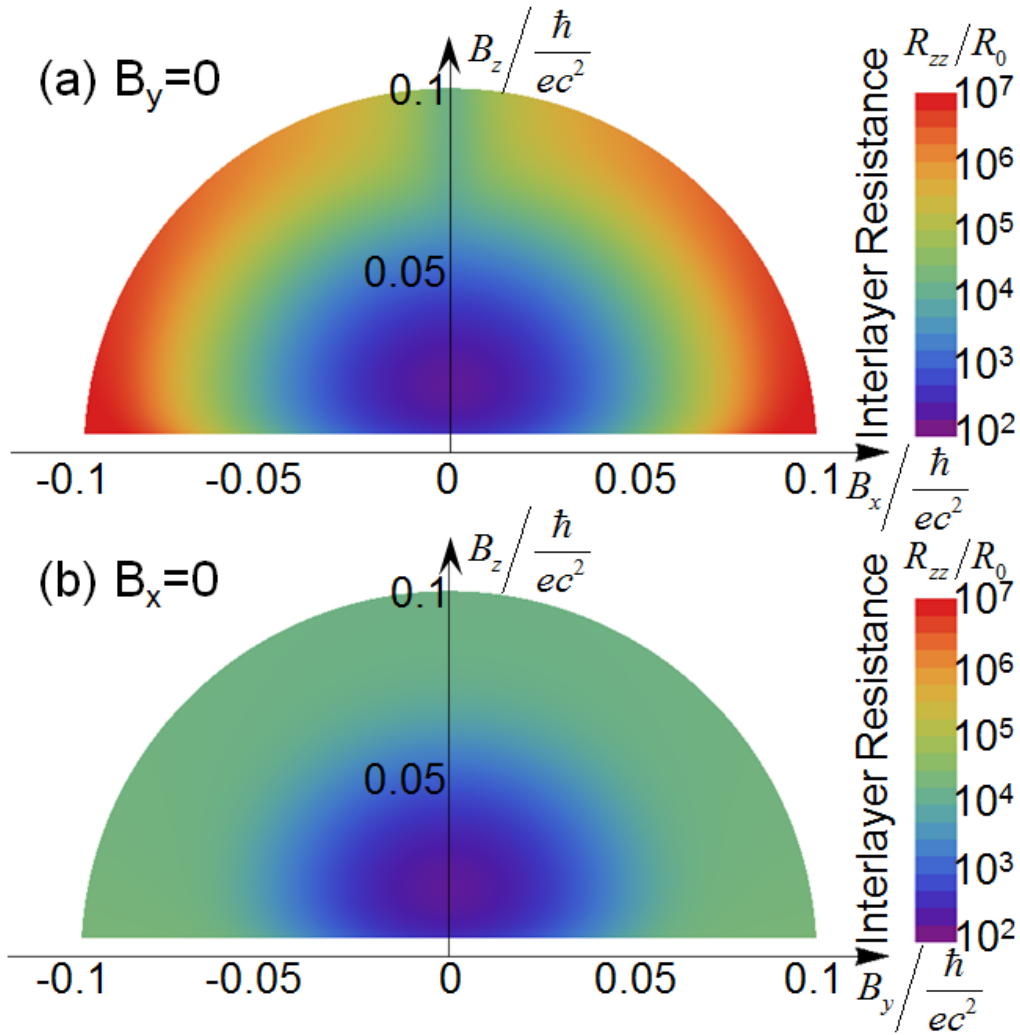

**Fig. 4.** (color)

Interlayer resistance $R_{zz}$ of the multilayer QH ferromagnet under the tilted magnetic fields. The orientation and the distance from the origin indicates the field orientation and strength, respectively. (a) The case when the field is tilted in the $xz$-plane perpendicular to the surface. (b) The case when the field is tilted in the $yz$-plane parallel to the surface. The same parameter set as Fig. 3 were employed.